\documentclass[11pt,a4paper]{article}
\usepackage[utf8]{inputenc}
\usepackage{amsmath}
\usepackage{amsfonts}
\usepackage{amssymb}
\usepackage{amsthm}
\usepackage{mathtools}
\usepackage[pdftex]{graphicx}
\usepackage{stmaryrd}
\usepackage[authoryear,comma,sectionbib, sort]{natbib} 
\usepackage{float}
\usepackage{mathrsfs}
\usepackage{bbold}
\usepackage{caption}
\usepackage{BOONDOX-cal}
\usepackage{graphics}
\usepackage[ruled,vlined]{algorithm2e}
\include{pythonlisting}
\usepackage{hyperref}

\usepackage{subfig}

\usepackage{tikz}
\usetikzlibrary{arrows,shapes}
\tikzstyle{var}=[draw=none,fill=none]
\tikzstyle{unobsvar}=[circle, draw,fill=none]
\tikzstyle{edge} = [draw,thick,->]
\tikzstyle{edge2} = [draw,thick, dashed,->]  
\usetikzlibrary{arrows,shapes}
\tikzstyle{edge2gris} = [draw,thick, dashed,->,CadetBlue]
\tikzstyle{edgebleu} = [draw,thick,->, NavyBlue]
\tikzstyle{edgerouge} = [draw,thick,->, OrangeRed]
\tikzstyle{edgeBleu} = [draw,thick,dotted,->, NavyBlue]
\tikzstyle{edgeRouge} = [draw,thick, dotted, ->, OrangeRed]
\tikzstyle{edgegris} = [draw,thick, ->, CadetBlue]

\renewcommand{\P}{{\rm I}\kern-0.14em{\rm P}}
\usepackage{float}

\newcommand{\R}{\mathbb{R}}

\newcommand{\1}{\mathbf{1}}
\newcommand{\Ind}{{\rm 1}\kern-0.26em{\rm I}}

\newcommand{\bM}{{\bf M}}

\newcommand{\bu}{{\bf u}}

\newcommand{\bw}{{\bf w}}

\newcommand{\bx}{{\bf x}}
\newcommand{\bX}{{\bf X}}

\newcommand{\by}{{\bf y}}

\newcommand{\bbeta}{{\boldsymbol \beta}}

\usepackage{dsfont}

\DeclareMathOperator{\argmin}{argmin}

\usepackage[utf8]{inputenc}

\renewcommand{\P}{\mathbb{P}}

\usepackage{color}

\newcommand{\blue}[1]{\textcolor{blue}{#1}}

\makeatletter
\newcommand*{\indep}{%
  \mathbin{%
    \mathpalette{\@indep}{}%
  }%
}
\newcommand*{\nindep}{%
  \mathbin{
    \mathpalette{\@indep}{\not}
  }%
}
\newcommand*{\@indep}[2]{%
  \sbox0{$#1\perp\m@th$}
  \sbox2{$#1=$}
  \sbox4{$#1\vcenter{}$}
  \rlap{\copy0}
  \dimen@=\dimexpr\ht2-\ht4-.2pt\relax
  \kern\dimen@
  {#2}%
  \kern\dimen@
  \copy0 
} 

\usepackage{xr}
\makeatletter
\newcommand*{\addFileDependency}[1]{
  \typeout{(#1)}
  \@addtofilelist{#1}
  \IfFileExists{#1}{}{\typeout{No file #1.}}
}
\makeatother

\usepackage{ulem}

\author{Nadim Ballout$^1$, Lola Étiévant$^2$, Vivian Viallon$^3$\footnote{Corresponding author: viallonv@iarc.fr}}
\date{%
    {\small $^1$ Univ Lyon, Univ Eiffel, IFSTTAR, Univ Lyon 1, UMRESTTE, UMR\_T9405, F-69500 Bron, France.\\%
    $^2$ Universit\'e Claude Bernard Lyon 1, CNRS UMR 5208, Institut Camille Jordan, Lyon, France.\\
    $^3$ Nutrition and Metabolism Branch, International Agency for Research on Cancer (IARC-WHO), Lyon, France;
    }
}
\title{{\bf On the use of cross-validation for the calibration of the adaptive lasso}}

\pdfoutput=1

\begin{document}

\maketitle

\section*{Abstract}
The adaptive lasso refers to a class of methods that use weighted versions of the $L_1$-norm penalty, with weights derived from an initial estimate of the parameter vector to be estimated. 
Irrespective of the method chosen to compute this initial estimate, the performance of the adaptive lasso critically depends on the value of a hyperparameter, which controls the magnitude of the weighted $L_1$-norm in the penalized criterion. As for other machine learning methods, cross-validation is very popular for the calibration of the adaptive lasso, that this for the selection of a data-driven optimal value of this hyperparameter. However, the most simple cross-validation scheme is not valid in this context, and a more elaborate one has to be employed to guarantee an optimal calibration. The discrepancy of the simple cross-validation scheme has been well documented in other contexts, but less so when it comes to the calibration of the adaptive lasso, and, therefore, many statistical analysts still overlook it. In this work, we recall appropriate cross-validation schemes for the calibration of the adaptive lasso, and illustrate the discrepancy of the simple scheme, using both synthetic and real-world examples. Our results clearly establish the suboptimality of the simple scheme, in terms of support recovery and prediction error, for several versions of the adaptive lasso, including the popular one-step lasso. 
\\
{\bf Keywords.} adaptive lasso, cross-validation, calibration, hyperparameter, tuning parameter, one-step lasso.

\section{Introduction}

High dimensional data are characterized by a number $p$ of variables larger, or at least not significantly lower, than sample size $n$. They have become ubiquitous in many fields, including biology, medicine, sociology, and economy \citep{giraud2014}. Their analysis raises a number of statistical challenges \citep{fanLi2006,hastie2009}, usually summarized under the term ``{\it curse of dimensionality}''. Consequently, it has attracted a lot of attention in the statistical literature over the past decades \citep{fanLi2006, hastie2009, donoho2000, hastie2015, buhlmannVan2011}. In particular, a variety of approaches based on the optimization of penalized versions of the log-likelihood have been developed, to estimate the true parameter vector $\bbeta^*=(\beta^*_1, \ldots, \beta^*_p)^T\in\R^p$ under high-dimensional parametric regression models \citep{tibshirani1996, huang2008asymptotic}. These approaches use a penalty term whose strength is controlled by a hyperparameter -- or tuning parameter --  and that is added to the loss-function so that the estimation can take advantage of some expected property of the true parameter vector $\bbeta^*$. For example, when $\bbeta^*$ is expected to be sparse, popular approaches rely on the use of $L_q$ penalties, for some $q\leq 1$. Among such approaches, the arguably most popular one is the lasso, which uses an $L_1$-norm penalty. Extensions such as the group lasso \citep{jacob2009group}, fused lasso \citep{tibshirani2005sparsity}, generalized fused lasso \citep{GenFused}, data shared lasso \citep{Gross, AutoRefLasso, NadimBinGraph, NadimSubTypes}, etc., rely on structured sparsity inducing norms, and can be used when some particular structured sparsity is expected in $\bbeta^*$. 

We will here focus on another extension of the lasso, namely the adaptive lasso \citep{zou2006, buhlmannVan2011}. It refers to a class of methods where the $L_1$-norm $\|\bbeta\|_1 = \sum_{j=1}^p |\beta_j|$ used in the standard lasso is replaced by a weighted version $\sum_{j=1}^p w_j |\beta_j|$. Weights $w_j\geq 0$ are typically data-driven, of the form $w_j= 1/(|\tilde \beta_j|+\varepsilon)$, with $\varepsilon\geq 0$ and $(\tilde \beta_j)_{1\leq j\leq p}$ some initial estimates of the parameters $(\beta^*_j)_{1\leq j\leq p}$. Three  popular versions of the adaptive lasso will be considered in this work: $(i)$ the original adaptive lasso introduced by \cite{zou2006}, where weights are derived from Ordinary Least Squares (OLS) estimates; $(ii)$ the version proposed by \cite{buhlmann2008}, where weights are computed from lasso estimates; and $(iii)$ the version proposed by \cite{zhang2008}, where weights are computed from ridge estimates.  In the rest of this article, the original adaptive lasso introduced by \cite{zou2006} will be referred to as the ols-adaptive lasso, the one proposed by \cite{zhang2008} as the ridge-adaptive lasso, while we will refer to the method described in \cite{buhlmann2008} as the one-step lasso, following their terminology. 

The adaptive lasso is very popular in practice. First, it can be implemented very easily and efficiently using algorithms originally developed for the lasso, such as the {\tt glmnet} R package \citep{glmnet}. Second, it has been shown to usually outperform the lasso. For example, in the fixed $p$ case, \cite{zou2006} established that the lasso estimates do not enjoy the asymptotic oracle property (in the sense of \cite{fanLi2006}), while the ols-adaptive lasso estimates do under mild conditions on the value of the hyperparameter that controls the strength of the weighted $L_1$-norm penalty. In addition, conditions ensuring support recovery in the non-asymptotic framework (which especially allows the study of the $p \gg n$ case) are weaker for the one-step lasso than for the lasso; see, e.g., Corollaries 7.8-7.9 and Section 2.8.3, in \cite{buhlmannVan2011}.  


As for other penalized approaches, the theoretical and empirical performance of the adaptive lasso critically depends on the value of the hyperparameter. Its optimal value involves unknown quantities, such as the variance of the noise under linear regression models, but also unknown constants related to the compatibility and irrepresentability conditions of the design matrix \citep{buhlmannVan2011}. Consequently, the practical selection of the tuning parameter, or calibration, also has attracted a lot of attention in the statistical literature \citep{chen2008extended, Chichignoud2016, giacobino2017quantile,arlot2019minimal}. Cross-validation is among the most commonly used strategies for the calibration of the adaptive lasso, and more generally of many statistical and machine learning methods;  see, e.g., Chapter 7 in \cite{hastie2009}. See also \cite{arlot2019Vpen, arlot2010survey}. Several versions of cross-validation have been described in the literature, but the $K$-fold version is arguably the most popular one in practice \citep{hastie2009}.

Briefly, $K$-fold cross-validation first consists in partitioning the original sample $D = (y_i, \bx_i)_{1\leq i\leq n}$ into $K\geq 2$ balanced folds $D^{(1)}, \ldots, D^{(K)}$, with $D = \cup_{k=1}^K D^{(k)}$. Then, it consists of $K$ steps, where at each step $k$, $(i)$ the fold $D^{(k)}$ is used as an ``independent'' test sample, while the remaining $K-1$ folds $D\setminus D^{(k)}$ are combined and jointly used as the training sample, $(ii)$  the estimator $\hat \bbeta_k$ is computed on the training sample $D\setminus D^{(k)}$, and $(iii)$ its prediction error, say ${\rm Pred.Err}(D^{(k)}, \hat \bbeta_k)$, is evaluated on the test sample $D^{(k)}$. The cross-validated prediction error is finally defined as the average of these $K$ prediction errors: $(1/K) \times \sum_k{\rm Pred.Err}(D^{(k)}, \hat \bbeta_k)$. We refer to \cite{bates2021crossvalidation} for additional details on the estimation of prediction error using cross-validation. This cross-validated prediction error can be used to assess the predictive performance of one estimator, or to compare the predictive performance among a set of estimators, or to select the tuning parameter by considering a set of estimators constructed with different values of the tuning parameter \citep{arlot2010survey}. However, for the cross-validation to be valid, it is crucial that the full learning algorithm is run on the training sample only, or, in other words, that data from the test sample are not used during the learning stage at all. For example, it is well known that if some variable selection is involved in the learning algorithm under study, this step has to be performed on the training sample only, and not on the full original sample \citep{hastie2009}. 

The adaptive lasso algorithm is made of two steps, where the first one consists in computing the initial estimates $(\tilde \beta_j)_{1\leq j\leq p}$ (through OLS estimation, the lasso, or ridge for example), and the second one consists in computing the adaptive lasso estimate {\it per se} with weights derived from these initial estimates. From the principle of the $K$-fold cross-validation recalled above, both steps have to be performed on the training sample to guarantee the validity of the calibration of the adaptive lasso \citep{StoneMervyn, krstajic2014cross}. Although this proper cross-validation scheme  was described in the literature \citep{kramer2009regularized, he2019improved}, the discrepancy of a more naive cross-validation scheme, where initial estimates are first derived from the full sample, and then the cross-validation is applied to select the hyperparameter of the adaptive lasso (focusing on the second step only), has not been well described and documented in the literature. In particular, \cite{kramer2009regularized} do not mention this simple scheme at all, while \cite{he2019improved} simply note that the it leads to possibly large false discovery rates, without stressing that this scheme is fundamentally inappropriate in this case. Likewise, \cite{buhlmann2008} recommend the use of cross-validation for the calibration of the tuning parameters in their one-step lasso for both the initial and final estimators (which is fine if a proper cross-validation scheme is used), but they do not stress the inappropriateness of the simple scheme for calibrating the final estimator, nor do they describe an appropriate scheme. Finally, except for a few exceptions such as the {\tt adalasso} function of the {\tt parcor}  \citep{kramer2009regularized} and the {\tt cvlasso} function of the  {\tt lassopack} package in STATA, proper implementations of the cross-validation for the adaptive lasso are lacking in publicly available adaptive lasso solvers. In particular, when using the  popular {\tt glmnet} R package \citep{glmnet}, it is tempting to implement the simple, and in this case improper, scheme, by simply using the {\tt cv.glmnet} function to select the hyperparameter of the adaptive lasso after computing the initial estimates (step 1) on the full sample. This improper cross-validation scheme, was used in the illustrative example of Section 2.8.1 of \cite{buhlmannVan2011}, and can also be found in the original version of the {\tt adapt4pv} R package \citep{courtois2021new}. Consequently, many statistical analysts still seem to apply this improper cross-validation scheme for calibrating the adaptive lasso  \citep{chang2020hdmac, pollard2021electrocardiogram, dessie2021novel}. As will be illustrated in Sections \ref{sec:Methods} and \ref{sec:Solution}, this generally leads to sub-optimal performance in terms of both support recovery and prediction accuracy. 



The rest of the article is organized as follows. In Section \ref{sec:Methods}, we start with a brief overview of the principles of the adaptive lasso, focusing on linear regression models for simplicity. Then, we illustrate the flaw of the simple $K$-fold cross-validation scheme when used for calibrating the adaptive lasso. A proper cross-validation scheme in this context is recalled and detailed in Section \ref{sec:Solution}. In Section \ref{sec:Simul}, we present results from a comprehensive simulation study where we empirically illustrate the suboptimality of the simple cross-validation scheme, in terms of support recovery and prediction accuracy. In Section \ref{sec:Appli}, we present results from an application on single-cell data. Concluding remarks are given in Section \ref{sec:Discussion}. 


\section{The adaptive lasso under the linear regression model}\label{sec:Methods}

\subsection{Main notation and working model}
As above, we will denote the sample size and the number of covariates by $n$ and $p$, respectively. For simplicity, we will focus on linear regression models of the form 
\begin{equation}
\by=\bX\bbeta^*+\xi, 
\label{lm}
\end{equation}
where $\by=(y_1, \ldots, y_n)^T \in \R^{n}$ is the response vector, $\bX=(\bx_{1},...,\bx_{n})^T \in \R^{n\times p}$ is the design matrix, $\bbeta^*=(\beta^*_1,...,\beta^*_p)^T\in \R^{p}$ is the $p$-dimensional vector of unknown parameters to be estimated, and  $\xi\in\R^n$ is some random noise. We further denote the support of $\bbeta^*$ by $J = \{j : \beta_j^*\neq 0\}$. 


For any positive integer $d\geq 1$, and any vector $\bu\in\R^d$, we will denote the usual Euclidian norm (or $L_2$-norm)  by $\|\bu\|_2$. We will let  ${\bf 0}_d$ and ${\bf 1}_d$ be the vectors of size $d$ with components all equal to 0 and 1 respectively, and  ${\bf I}_d$ be the $d\times d$ identity matrix. For any real matrix $\bM = (M_1, \ldots, M_d)\in \R^{n\times d}$, and any subset $E\subseteq\{1, \ldots, d\}$, $\bM_E$ will denote the submatrix composed of the columns $(M_j)_{j\in E}$. We will further denote the cardinality of $E$ by $|E|$.

Finally, for any sample $D_0=\{y_i,\bx_i\}_{i\in I_0}$, with $I_0$ a given set of integers, and any estimator $\hat\bbeta$ of  $\bbeta^*$, we will denote by $$ {\rm Pred.Error} (D_0, \hat\bbeta) = \frac{1}{|I_0|} \sum_{i \in I_0} (y_i - \bx_i^T \hat \bbeta)^2 $$ the prediction error corresponding to $\hat \bbeta$ evaluated on the sample $D_0$. 

\subsection{The lasso and adaptive lasso}

For any $\lambda \geq 0$, the lasso estimator $\hat{\bbeta}_{\rm lasso}(\lambda)$ \citep{tibshirani1996} is defined as any minimizer over $\bbeta\in\R^p$ of the penalized criterion
\begin{equation}
\frac{||\by - \bX\bbeta||_2^2}{n} + \lambda \sum_{j=1}^{p} |\beta_j|.
\label{lasso}
\end{equation} 
The tuning parameter $\lambda$ controls the amount of regularization through the $L_1$-norm $\|\bbeta\|_1 = \sum_{j=1}^{p} |\beta_j|$. In practice, an appropriate value for this parameter is needed to guarantee good statistical performance for $\hat{\bbeta}_{\rm lasso}(\lambda)$, with respect to both support recovery and prediction accuracy. As mentioned above, a popular strategy relies on $K$-fold cross-validation \citep{hastie2009} whose pseudo-code is recalled in Algorithm \ref{alg:CV} in Appendix \ref{Appendix:CV}. Let $\lambda^{\rm CV}$ be the value of $\lambda$ selected by $K$-fold cross-validation, and let $\hat{\bbeta}_{\rm lasso}^{\rm CV}=\hat{\bbeta}_{\rm lasso}(\lambda^{\rm CV})$ denote one particular solution of \eqref{lasso} with $\lambda$ set to $\lambda^{\rm CV}$. 

Now, denote by $\bw = (w_1, \ldots, w_p) \in \R^p_{\geq 0}$ a given vector of non-negative weights. For any $\lambda\geq 0$, the adaptive lasso estimator $\hat{\bbeta}_{\rm ada}(\lambda; \bw)$ is defined  \citep{zou2006} as any minimizer over $\bbeta\in\R^p$ of the criterion
\begin{equation}
\frac{||\by - \bX\bbeta||_2^2}{n} + \lambda\sum_{j=1}^{p} w_j|\beta_j|.
\label{adlasso}
\end{equation}
The adaptive lasso reduces to the standard lasso for the particular choice of the weight vector $\bw = \1_p$: any solution $\hat{\bbeta}_{\rm lasso}(\lambda)$ is also a solution $\hat{\bbeta}_{\rm ada}(\lambda; \1_p)$. In practice, weights are usually set to $w_j=1/|\tilde{\beta}_j|$, or $w_j=1/(|\tilde{\beta}_j|+\varepsilon)$, with     $\tilde{\bbeta}=(\tilde{\beta}_1,...,\tilde{\beta}_p)^T$ an initial estimator of $\bbeta^*$, and $\varepsilon$ some non-negative real number. A positive value for the $\varepsilon$ parameter guarantees that every component of the weight vector is finite, so that every component of $\hat{\bbeta}_{\rm ada}(\lambda; \bw)$ has a chance to be non-zero. If the initial estimator is good enough, then $\tilde{\beta}_j$ is close to $0$ for $j\notin J$, and less so for $j\in J$: if, in addition, $\varepsilon$ is null or close enough to 0, weights $w_j$ are large for $j\notin J$, and less so for $j\in J$. Then, components $j\notin J$ of the parameter vector are more heavily penalized than components $j\in J$. Several initial estimates can be used to derive the weight vector $\bw$. When $p<n$, \cite{zou2006} suggests the use of $\bw_{\rm OLS} = 1/|\tilde \bbeta_{\rm OLS}|$, where $\tilde \bbeta_{\rm OLS} = (\bX^T\bX)^{-1}\bX^T \by$ is the OLS estimator. Under mild conditions, and considering the fixed $p$ case, \cite{zou2006} established the asymptotic oracular property for this ols-adaptive lasso: specifically, the ols-adaptive lasso is consistent in terms of variable selection (or sparsistent), and the distribution of $(\hat{\beta}_{\rm adap,j}(\lambda; \bw_{\rm OLS}))_{j\in J}$ is Gaussian with the same expectation and covariance matrix than that of $(\bX_J^T\bX_J)^{-1}\bX_J^T \by$, the OLS estimator that would be obtained if $J$ were known in advance. The ols-adaptive lasso inherits these good properties from the $\sqrt{n}$-consistency of the OLS estimate $\tilde \bbeta_{\rm OLS}$ in  low-dimensional settings. However, the OLS estimate is less attractive when $p>n$. In these high-dimensional settings, \cite{zhang2008} proposed the ridge-adaptive lasso, where ridge regression is used for the computation of the weights: $\bw_{{\rm ridge}} = 1/|\hat\bbeta_{\rm ridge}^{\rm CV}|$. More precisely, $\hat\bbeta_{\rm ridge}^{\rm CV}$ denotes the estimator produced by a ridge regression with tuning parameter selected by $K$-fold cross-validation, and we recall that the ridge regression is a penalized approach similar to the lasso, but with the $L_1$-norm in the penalty replaced by the squared $L_2$-norm $\|\bbeta\|_2
^2$ \citep{hoerl1970ridge}.  Alternatively, \cite{buhlmann2008} suggest the use of weights $\bw_{{\rm 1-step}}$ derived from $\hat{\bbeta}_{\rm lasso}^{\rm CV}$. This leads to what they refer to as the one-step lasso. Thanks to the so-called screening property of $\hat{\bbeta}_{\rm lasso}^{\rm CV}$, the one-step lasso  was shown to be sparsistent under weaker irrepresentability conditions than those required for the lasso \citep{buhlmannVan2011}. Other choices for the weights have been proposed in the literature: for example, \cite{huang2008} suggested the use of univariate OLS estimators.

\subsection{Illustration of the defect of the simple $K$-fold cross-validation scheme for the calibration of the adaptive lasso} \label{sec:CV}


As recalled in the Introduction, the cross-validated prediction error $(1/K) \times \sum_k{\rm Pred.Err}(D^{(k)}, \hat \bbeta_k)$ is widely used to calibrate penalized approaches, including the adaptive lasso. However, many statistical analysts still seem to implement too simple of a scheme for the calibration of the adaptive lasso \citep{chang2020hdmac, pollard2021electrocardiogram, dessie2021novel, courtois2021new}. More precisely, they consider the weight vector $\bw\in \R^p_{\geq 0}$ derived from initial estimates computed on the full original sample as a ``given'' (fixed) weight vector. Then, they improperly use the simple cross-validation scheme to compare the predictive performance of the set of estimators $(\hat{\bbeta}_{\rm ada}(\lambda_r; \bw))_{1\leq r \leq R}$, for any given sequence $\Lambda = (\lambda_1, \ldots, \lambda_R)$ of candidate values for the hyperparameter, and finally select the optimal hyperparameter value, say, $\lambda^{\rm CV}(\bw)$, and the corresponding optimal estimator $\hat{\bbeta}_{\rm ada}(\lambda
^{\rm CV}(\bw); \bw)$. See the pseudo-code given in Algorithm \ref{alg:CV} in Appendix \ref{Appendix:CV} for the detailed description of this simple, and  in this case improper, cross-validation scheme.


Below, we present results from a first simulation study, whose main objectives are to illustrate $(i)$ the good performance of the simple scheme when used for the calibration of the standard lasso, and $(ii)$ its poor performance when applied for the calibration of adaptive lasso.  Results from a more comprehensive simulation study will be presented in Section \ref{sec:Simul}. 

In this first synthetic example, we generate one sample $D = (y_i, \bx_i)_{1\leq i\leq n}$, made of $n=1,000$ observations under the linear regression model \eqref{lm} with $p = 1,000$. We first set $\bbeta^*=(\beta^*_1, \ldots, \beta^*_p)^T$ with $\beta^*_j=0$ for all $j\geq 11$, and $\beta^*_j = \iota_j 0.5$ for all $j\leq 10$, where $\iota_j$ is a $\{-1, 1\}$-binary random variable, with $\P(\iota_j = 1)=1/2$. Then, for each $i=1, \ldots, n$, we generate a Gaussian random noise $\xi_i\sim N(0,1)$,  a Gaussian vector of covariates $\bx_i = (x_{i,1},...,x_{i,p}) \sim N({\bf 0}_p,{\bf I}_{p})$, and finally the outcome $y_i = \bx_i^T \bbeta^* +\xi_i$. Similarly, we generate one independent test sample ${\cal D}=(y_i, \bx_i)_{n+1\leq i\leq n+N}$, made of $N=10,000$ observations drawn under the same linear model. Then, for any given weight vector $\bw$, the {\tt glmnet} R package is used to compute the (adaptive) lasso estimator on an appropriate sequence $\Lambda= \Lambda(D, \bw) = (\lambda_1(D, \bw), \ldots, \lambda_{100}(D, \bw))$ of 100 decreasing values for the tuning parameter. The {\tt glmnet} R package is used to compute the 10-fold cross-validated prediction error of $\hat{\bbeta}_{\rm ada}(\lambda_r(D, \bw); \bw)$, for $r=1, \ldots, 100$. For comparison, the true prediction error of each estimator is estimated on the independent test sample ${\cal D}$. 

\begin{figure}[t]
\includegraphics[width=\textwidth]{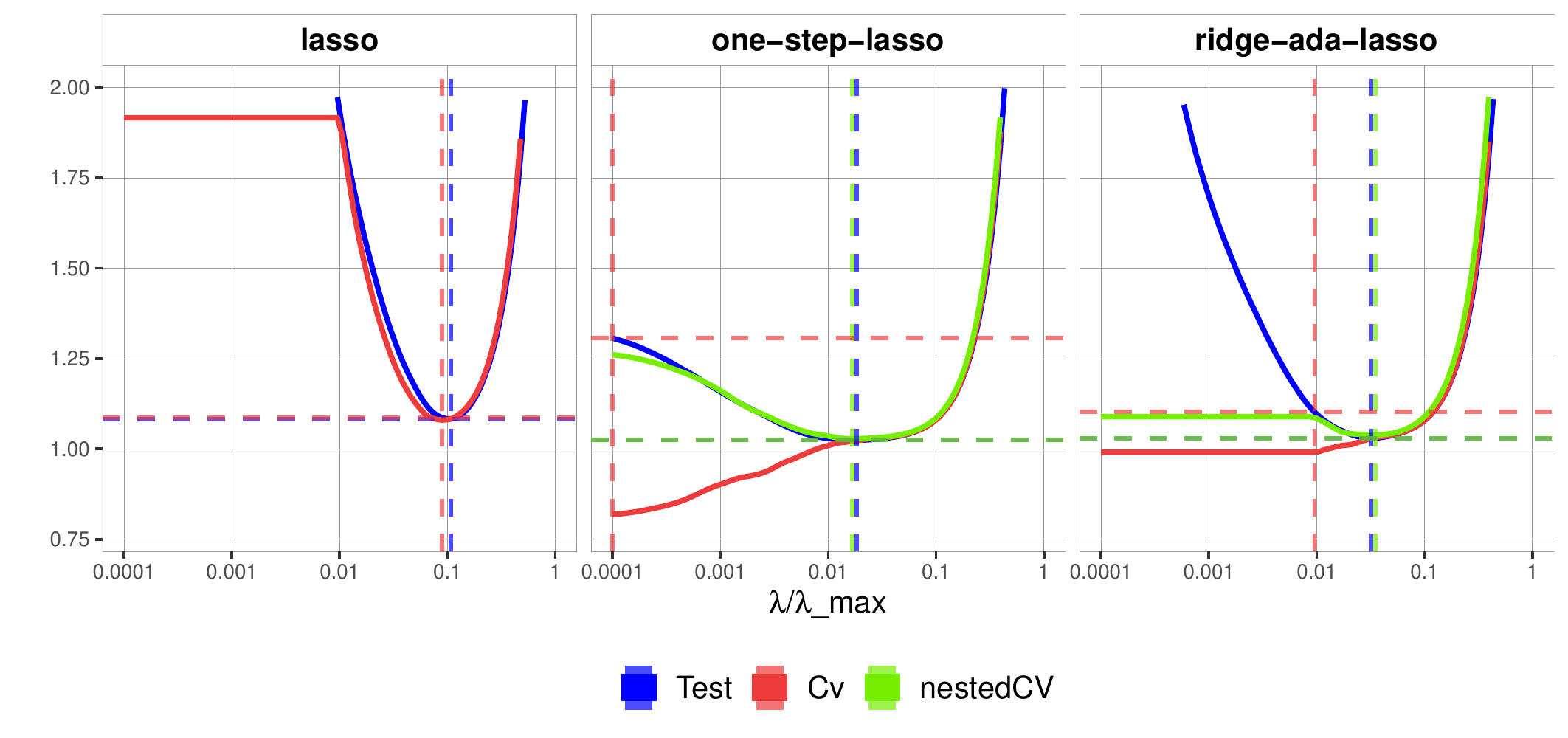}
\caption{Comparison between the prediction error estimated on an independent test sample (in blue) and the prediction error estimated via the simple cross-validation scheme (in red), for the lasso (left), the one-step-lasso (middle), and the ridge-adaptive lasso (right). For the latter two, the cross-validated prediction error estimated using a proper nested cross-validation scheme is also presented (in green). Vertical dotted lines represent the value of the tuning parameter for which each particular curve is minimized (red: cross-validated prediction error using the simple scheme; green: cross-validated prediction error using the nested scheme; blue: prediction error estimated on an independent test sample). Horizontal dotted lines represent the value of the estimated true prediction error for these particular values of the tuning parameter.}\label{fig1}
\end{figure}

Given the relatively high-dimensional setting of this first simulation study, the ols-adaptive lasso is not considered here. Figure \ref{fig1} presents the results for the lasso ($\bw = \1_p$),  the one-step lasso ($\bw = \bw_{\rm 1-step} = 1/|\hat{\bbeta}_{\rm lasso}^{\rm CV}|$), and the ridge-adaptive lasso ($\bw = 1/|\hat{\bbeta}_{\rm ridge}^{\rm CV}|$). For the latter two, we also report the cross-validated prediction error using a proper nested cross-validation scheme, which we will describe in more details in the next Section. For comparability, the $x$-axis corresponds to the tuning parameter sequence represented as a fraction of the data-specific maximal value $\lambda_1(D, \bw)$. 


First consider the standard lasso (left panel). In this case, the cross-validated prediction error using the simple cross-validation scheme does a fairly good job in approximating the "true" prediction error (estimated on the test sample) on a wide range of $\lambda$-values. In particular, the $\lambda$-value at which the cross-validated prediction error is minimized (vertical dotted red line) is very close to that at which the prediction error on the test sample is minimized (vertical dotted blue line). Moreover, the prediction errors evaluated on the test sample at these two $\lambda$-values (horizontal dotted red and blue lines, respectively) are indistinguishable: in this example, the lasso estimate $\hat{\bbeta}_{\rm lasso}^{\rm CV}$ selected by cross-validation is therefore nearly optimal with respect to prediction error. However, the simple cross-validation scheme does not perform that well for the two versions of the adaptive lasso. In particular, when using the simple scheme for the one-step lasso, the cross-validated prediction error constantly decreases as the tuning-parameter decreases, a behavior that we observed on many other simulation designs as well (results not shown). Then, the $\lambda$-value at which the cross-validated prediction error is minimized (vertical dotted red line) is very different from the minimizer of the prediction error evaluated on the test sample. In this example, the support of the one-step lasso estimator is too large if the tuning parameter is selected via the simple cross-validation scheme. This will also usually be the case in the more comprehensive simulation study presented in Section \ref{sec:Simul}. Moreover, the prediction errors evaluated on the test sample at  these two $\lambda$-values (horizontal dotted red and blue lines, respectively) differ substantially: the one-step-lasso estimator calibrated using the simple cross-validation scheme is far from optimal with respect to prediction error in this example. A similar albeit less pronounced behavior is observed in the case of the ridge-adaptive lasso, confirming that the simple $K$-fold cross-validation scheme is not valid for the calibration of the adaptive lasso. Conversely, these defects disappear when using a proper scheme whose principle will be recalled in details in the next Section \citep{kramer2009regularized, he2019improved, ahrens2020lassopack}, which is based on nested cross-validation scheme \citep{varma2006bias, krstajic2014cross}. Using this proper scheme, the $\lambda$-value at which the corresponding cross-validated prediction error is minimized (vertical dotted green line) is close to the minimizer of the true prediction error estimated on the test sample for both the one-step lasso and ridge-adaptive lasso. Moreover, the prediction errors evaluated on the test sample at these two $\lambda$-values (horizontal dotted green line) are very close to the optimal prediction error: in this example, the one-step-lasso and the ridge-adaptive lasso estimators calibrated using the proper nested cross-validation scheme are nearly optimal with respect to prediction error.

\section{Proper cross-validation schemes for the adaptive lasso}\label{sec:Solution}

In this Section, we provide details on the rationale and principle of proper cross-validation schemes for the adaptive lasso, as previously introduced in \cite{kramer2009regularized}, \cite{he2019improved} and \cite{ahrens2020lassopack}.

\subsection{Additional notation}

For any sample of observations $D_0=\{y_i,\bx_i\}_{i\in I_0}$, with $I_0$ a given set of integers, any vector of non-negative weights $\bw$, and any non-negative $\lambda$ value, we let Lasso$(D_0, \bw, \lambda)$ denote one particular solution of the adaptive lasso \eqref{adlasso}, when computed on sample $D_0$ with weights $\bw$ and tuning parameter $\lambda$. For any positive integer $K\geq 2$, let sCVLasso$(D_0,\bw, K)$ be an adaptive lasso estimator computed on $D_0$ with weights $\bw$ and the tuning parameter set to its optimal value according to the simple $K$-fold cross-validation scheme of Algorithm \ref{alg:CV} in Appendix \ref{Appendix:CV}; this scheme is improper unless the weights $\bw$ are independent from $D_0$. On the other hand, nestedCVLasso$(D_0,\bw, K)$ will denote an adaptive lasso estimator computed on $D_0$ with weights $\bw$ and the tuning parameter set to its optimal value according to an appropriate $K$-fold cross-validation scheme. This proper scheme will be described in details below. We use the notation nestedCVLasso to emphasize that this cross-validation scheme relies on two nested cross-validations in the case of the one-step lasso and the ridge-adaptive lasso; however, this is not the case for the ols-adaptive lasso, and our notation  ``nestedCVLasso" is a bit misleading in the case of the ols-adaptive-lasso. On the other hand, we will use the shorthand sCVRidge$(D_0,K)$ to denote the ridge estimator computed on $D_0$ with a tuning parameter set to its optimal value according to the simple $K$-fold cross-validation, and that will be used to compute the weights in the ridge-adaptive lasso.  

\subsection{Rationale and detailed presentation}

We start by recalling the rationale for the defect of the simple $K$-fold cross-validation when used for the calibration of the adaptive lasso described and illustrated in Section \ref{sec:CV}. As mentioned in the Introduction, the principle of cross-validation is to mimic ``independent'' test samples. Then, for the cross-validation to be valid, at each step $k$ of the $K$-fold cross-validation, the whole estimation procedure should be performed on the training sample $D\setminus D^{(k)}$, and should not use any information from the test sample $D^{(k)}$. However, the weights used in the adaptive lasso are derived from initial estimates computed on the entire original sample $D$. Therefore, when using the simple cross-validation scheme of Algorithm \ref{alg:CV} in Appendix \ref{Appendix:CV} for the calibration of the adaptive lasso, and considering the estimation of the adaptive lasso estimator as a whole, data from the test samples are used in the first step of the adaptive lasso, which makes the simple cross-validation scheme not valid in this particular framework. 

A proper cross-validation scheme for the calibration of the adaptive lasso estimator is described in Algorithm \ref{alg:1step} \citep{kramer2009regularized, he2019improved, ahrens2020lassopack}. The main difference with the simple scheme lies in Step 2. Under the simple scheme, this step consists of Step 2-a, which relies on sCVLasso$(D,\bw, K)$, while it is replaced by Step 2-a' in the proper scheme, which relies on nestedCVLasso$(D,\bw, K)$, detailed in Algorithm \ref{alg:NCV} in the particular case of the one-step lasso. The key difference between sCVLasso (see Algorithm \ref{alg:CV} in Appendix \ref{Appendix:CV}) and  nestedCVLasso is highlighted in blue in Algorithm \ref{alg:NCV}: at each step $k$ of the proper cross-validation scheme, weights used for the adaptive lasso are first recomputed on the training sample $D\setminus D^{(k)}$, so that the whole estimation of the adaptive lasso estimator uses information from the training sample only, before computing the corresponding prediction error on the test sample $D^{(k)}$. In the case of the one-step lasso (or the ridge-adaptive lasso), this leads to a nested cross-validation scheme \citep{varma2006bias, krstajic2014cross}. More precisely, an appropriate sequence $\Lambda$ of candidate $\lambda$ values for the hyperparameter has first to be chosen, which typically depends on the weights computed on the whole original sample (we briefly get back to this point below). Then, at each step $k$ of the ``outer'' $K$-fold cross-validation, one ``inner'' standard cross-validation is performed to compute the optimal lasso (or ridge) estimator on the training sample $D\setminus D^{(k)}$, from which the weights are derived, before the corresponding adaptive lasso estimator is computed for each of the $\lambda$ values of the sequence, and their predictive performance is eventually evaluated on the test sample $D^{(k)}$. For each $\lambda$ value of the sequence, the predictive performance is then averaged over the $K$ folds. The optimal value for the hyperparameter is defined as the value that minimizes this averaged criterion. The optimal adaptive lasso estimator finally corresponds to the adaptive lasso computed on $D$, with weights $\bw$ also computed on $D$, and the hyperparameter set to this optimal value. 

A first remark is that, in the case of the ols-adaptive lasso, the proper cross-validation scheme still involves the computation of weights on each training sample in the ``outer'' loop; however, because weights are derived from OLS estimation, no hyperparameter has to be selected for their computation, hence no inner loop is required. In other words, for the ols-adaptive lasso, the proper cross-validation scheme does not involve nested loops. A second important remark concerns the sequence of candidate values for the hyperparameter. Above, we propose to use the data-driven sequence $\Lambda:= \Lambda(D, \bw) = (\lambda_1, \ldots, \lambda_{R})$ computed from the entire original sample, with weights $\bw$ also computed on the entire original sample. It could be argued that, by doing so, the estimation process is not completely independent from the test samples. We tested alternative strategies for the selection of the sequence $\Lambda$ in our simulation, including the data-driven sequence $\Lambda(D, \1_p)$ computed from the full original sample with weight vector set to $\1_p$ that is used in the {\tt adalasso} function of the {\tt parcor} R package \citep{kramer2009regularized}, but the results were virtually unchanged.   

\begin{algorithm}[H]\label{alg:1step}
\SetAlgoLined
\KwData{Sample: $D=\{y_i,\bx_i\}_{i=1}^{n}$, Version of the adaptive lasso, Number of folds: $K$, and parameter $\varepsilon\geq0$}
\KwResult{$\hat{\bbeta}_{\rm ada}^{\rm CV}$}
{\bf Step 1: Computation of the initial estimates and weights}\;
 (1-a) {\it Initial estimates: either (i), (ii) or (iii) below, depending on the considered version of the adaptive lasso} \;
  \hspace{1cm}(i): \, $\tilde{\bbeta} = {\rm sCVLasso}(D, \1_p, K)$   \!\,  \tcc{one-step lasso}
  \hspace{1cm}(ii): \ $\tilde{\bbeta} = {\rm OLS}(D)$  \hspace{1cm} \!\, \, \, \, \, \tcc{ols-adaptive lasso}
  \hspace{1cm}(iii): $\tilde{\bbeta} = {\rm sCVRidge}(D, K)$  \, \, \, \tcc{ridge-adaptive lasso}
 (1-b) {\it Weights:} $\bw = 1/(|\tilde{\bbeta}|+\varepsilon)$\;
{\bf Step 2: Computation of the final estimates}\; 
 (2-a) \ $\hat \bbeta_{\rm ada}^{\rm CV}(\bw) = {\rm sCVLasso}(D, \bw, K)$\; 
 \blue{(2-a') $\hat{\bbeta}_{\rm ada}^{\rm CV}(\bw)={\rm nestedCVLasso}(D, \bw, K)$\;}
 \caption{Cross-validation for the adaptive lasso. Versions $(i)$, $(ii)$ and $(iii)$ of step (1-a) correspond to the one-step lasso, the ols-adaptive Lasso, and the ridge-adaptive lasso, respectively. The simple (and improper) scheme corresponds to the algorithm ran with Step (2-a), while Step (2-a') (detailed in Algorithm \ref{alg:NCV}) should be used instead. }
\end{algorithm}

\begin{algorithm}[H]\label{alg:NCV}
\SetAlgoLined
\KwData{Sample: $D=\{y_i,\bx_i\}_{i=1}^{n}$, Weights: $\bw$, Number of folds: $K$, and parameter $\varepsilon\geq0$}
\KwResult{$\hat{\bbeta}_{\rm ada}^{nCV}(\bw)=\hat{\bbeta}_{\rm ada}(\lambda^{nCV}, \bw, D)$}
 Computation of a sequence $\Lambda:= \Lambda(D, \bw) = (\lambda_1, \ldots, \lambda_{R})$\;
 Division of $D$ into $K$ folds: $D= \cup_{k=1}^K  D^{(k)}$\;
 \For{$k\in\{1, \ldots, K\}$}{
 \blue{\tcc{Computation of the weights on $D\setminus D^{(k)}$}}
 \hspace{1cm} \blue{$\tilde{\bbeta}^{(k)} = {\rm sCVLasso}(D\setminus D^{(k)}, \1_p, K)$}\;
 \hspace{1cm} \blue{$\bw_k = 1/(|\tilde{\bbeta}^{(k)}|+\varepsilon)$}\;
 \For{$r\in \{1, \ldots, R\}$}{
    $\hat \bbeta_{r,k}$ = Lasso$(D\setminus D^{(k)}, \blue{\bw_k}, \lambda_r)$\;
   $E_{r,k}$ = Pred.Error$(D^{(k)}, \hat \bbeta_{r,k})$\;
  }
 }
 $r^* = \argmin_r \big\{\sum_{k=1}^K E_{r,k}\big\}$\;
 $\lambda^{nCV} = \lambda_{r^*}$ \;
 $\hat{\bbeta}_{\rm ada}^{nCV}(\bw) =$ Lasso$(D, \bw, \lambda^{nCV})$\;
 \caption{Details on the nested $K$-fold cross-validation scheme (nestedCVLasso) for the calibration of the adaptive lasso, in the particular case of the one-step lasso. }
\end{algorithm}

\section{Simulation study}\label{sec:Simul}


We now present results from a more comprehensive simulation study, which extends the simple one presented in Section \ref{sec:CV}. As before, we generate a sample $D=(y_i, \bx_i)_{1\leq i\leq n}$, made of $n=500$ observations drawn under linear regression model \eqref{lm}. Here, we make the number of covariates $p$ vary in $\{100,500,1000\}$, and set the number of relevant covariates to $p_0=10$. We randomly select the support $J$ of $\bbeta^*$, with $|J|=p_0$. Then, we set $\beta_j^*=0$ for all $j\notin J$, and $\beta_j^* = \iota_j \beta$, for all $j \in J$, where $\iota_j$ is a $\{-1, 1\}$-binary variable, with $\P(\iota_j = 1)=1/2$. As for the signal strength $\beta$, we make it 
vary in $\{1/4, 1/2, 1, 3/2\}$. Then, for each $i=1, \ldots, n$, we generate a Gaussian random noise $\xi_i\sim N(0,1)$, a Gaussian vector of covariates $\bx_i = (x_{i,1},...,x_{i,p}) \sim N({\bf 0}_p,{\boldsymbol \Sigma})$, with ${\boldsymbol \Sigma}$ a symmetric $p\times p$ matrix, with element $(k,j)$ equal to $\sigma_{k,j} = 0.3^{|k-j|}$, for $1\leq k,j\leq p$. Finally the outcome for observation $i$ is generated as $y_i = \bx_i^T \bbeta^* +\xi_i$. Additionally, we generate one independent test sample ${\cal D}=(y_i, \bx_i)_{n+1\leq i\leq n+N}$, made of $N=10,000$ observations drawn under the same linear model. 

We consider the one-step lasso and the ridge-adaptive lasso as before. The ols-adaptive lasso is also considered in the low-dimensional scenario where  $p=100$. As in the previous study, weights were derived from initial estimates $\tilde \bbeta$ as $w_j = 1/|\tilde \beta_j|$; other choices for the parameter $\varepsilon \in\{0, 10^{-6}, 10^{-4}, 10^{-2}\}$ were tested, and led to very similar results (not shown). For each method, the optimal adaptive lasso estimator is computed following either the simple 10-fold cross-validation scheme (Algorithm \ref{alg:1step} with Step 2-a; we will refer to these estimators as the one-step lasso CV, the ridge-adaptive lasso CV, and the ols-adaptive lasso CV respectively) or the 10-fold ``nested'' cross-validation scheme (Algorithm \ref{alg:1step} with Step 2-a'; we will refer to these estimators as the one-step lasso nested CV, the ridge-adaptive lasso nested CV, and the ols-adaptive lasso nested CV, respectively; keep in mind that for the ols-adaptive lasso nested CV, the terminology is a bit misleading, as the cross-validation scheme does not involve nested loops). For comparison, results from the lasso estimator calibrated by standard 10-fold cross-validation (see Algorithm \ref{alg:CV} in Appendix \ref{Appendix:CV}) are presented; we will refer to this estimator as the lasso CV. We used functions from the {\tt glmnet} R package to compute these different estimators. We evaluate the (signed) support accuracy, the precision, the recall and the prediction error corresponding to each estimator $\hat \bbeta$. More precisely, the signed support accuracy is defined as ${\rm sACC} = \{\sum_{j=1}^p \Ind({\rm sign}(\beta^*_j) ={\rm sign}(\hat\beta_j)) \}/p$ , where $\Ind$ is the indicator function, and ${\rm sign}(\cdot)$ the sign function, that is ${\rm sign}(x) = +1$ if $x>0$, ${\rm sign}(x) =-1$ if $x<0$, and ${\rm sign}(x) =0$ if $x=0$. Precision and recall are defined as $\sum_{j=1}^p\Ind[\beta_j^*\neq0,\hat{\beta}_j\neq0]/\sum_{j=1}^p\Ind[\hat\beta_j\neq0]$  and $\sum_{j=1}^p\Ind[\beta_j^*\neq0,\hat{\beta}_j\neq0]/p_0 $,  respectively. As for the prediction error, we simply compute ${\rm Pred.Err}({\cal D}, \hat \bbeta)$, as before. These four criteria are averaged over the 100 replications we consider for each combination of values for the parameters $(p, \beta)$.  

\begin{figure}[t]
\includegraphics[width=\textwidth]{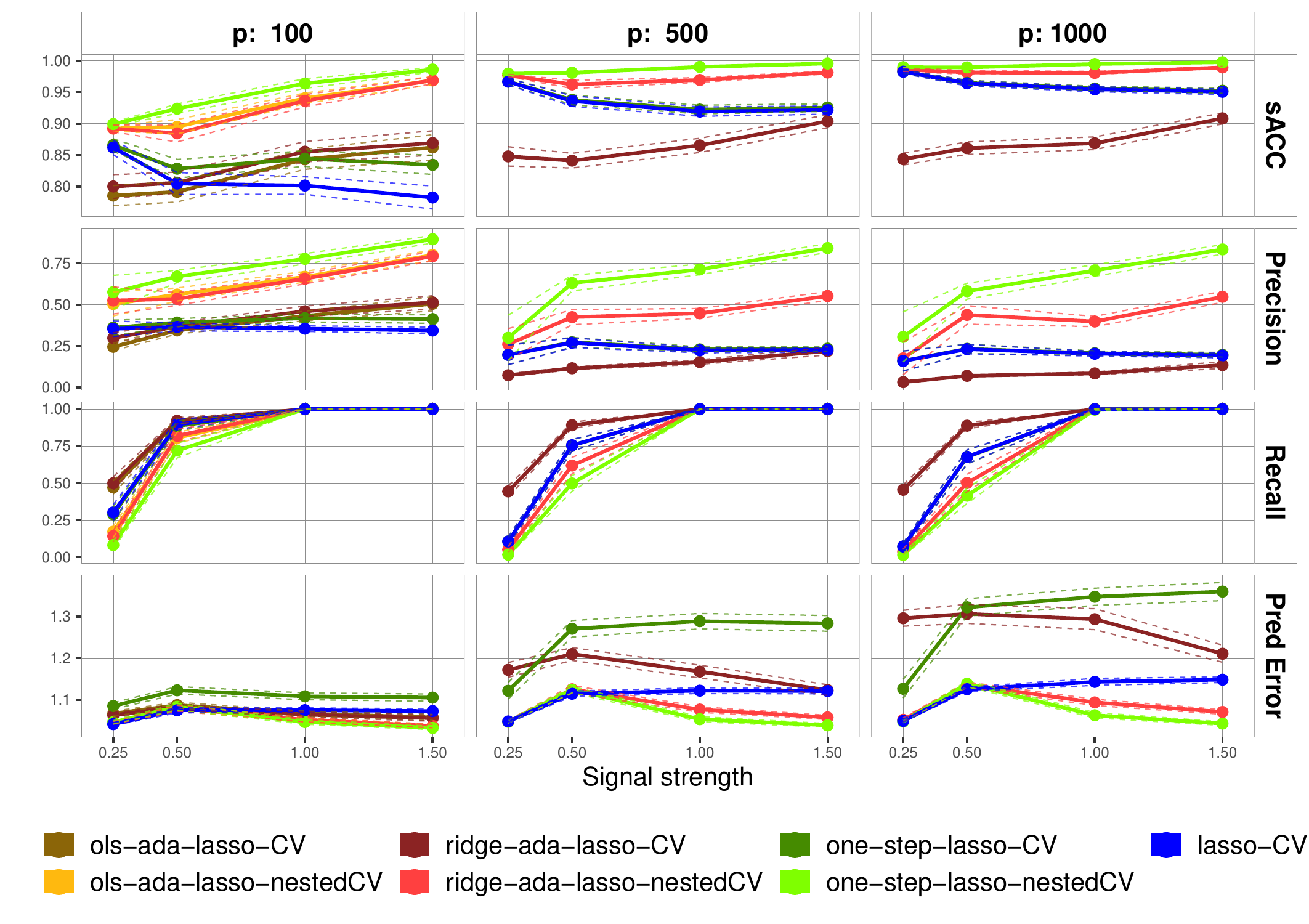}
\caption{Results of the simulation study. Solid lines represent the averaged criteria, while dashed lines represent the corresponding 95\% confidence intervals.}\label{SimRandomData-n500-All}
\end{figure}

Figure \ref{SimRandomData-n500-All} presents the results. First, focusing on the adaptive lasso estimators, our results illustrate that the proper cross-validation scheme yields better performance than the simple scheme, in terms of both prediction error and support accuracy. In terms of support recovery, the gain actually only concerns precision (or, equivalently, specificity). This observation confirms what we already observed in Section \ref{sec:CV}: the simple cross-validation scheme generally leads to the selection of too small of hyperparameters, hence poor precision/specificity. Conversely, recall (which is related to sensitivity) is usually better with the simple cross-validation scheme, but the balance is clearly in favour of the proper cross-validation scheme, as it usually yields substantially better overall support recovery. We also observe that, when correctly calibrated, the adaptive lasso usually outperforms the lasso in terms of support recovery (again, the gain comes from a better precision/specificity, as expected), and prediction error. However,  the two methods achieve very similar performance for low signal strength.

\begin{figure}[t]
\centering
\includegraphics[width=\textwidth]{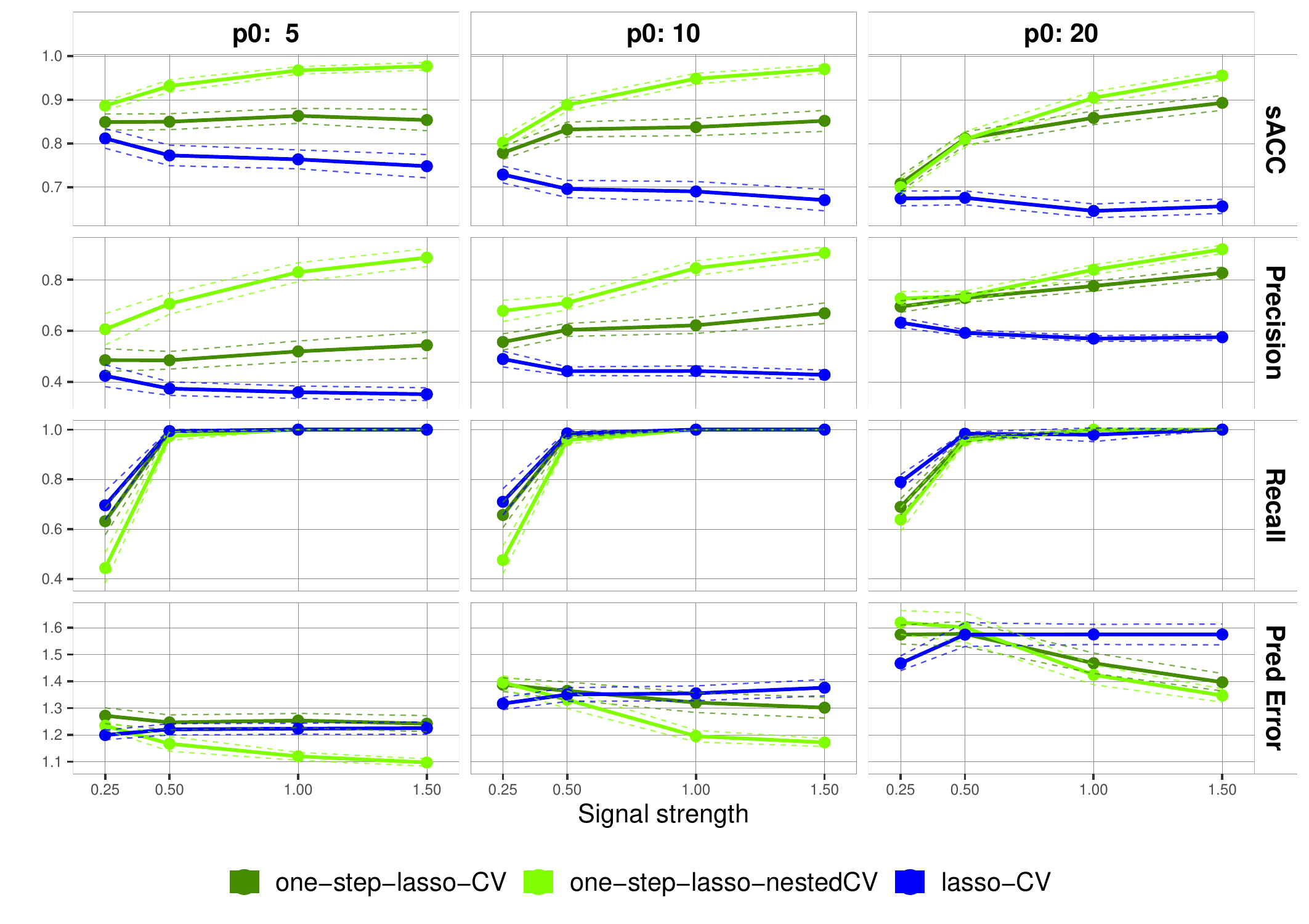}
\caption{Results of the additional simulation study inspired by the single cell data. Solid lines represent the averaged criteria, while dashed lines represent the corresponding 95\% confidence intervals.}\label{SimRealData-n100-Zoom}
\end{figure}

Finally, Figure \ref{SimRealData-n100-Zoom} presents results from an additional simulation study, using a more realistic scenario, where predictors are highly correlated. Specifically, our synthetic data are generated just as above, but this time with $n=100$ and $p=44$, and with the variance-covariance matrix ${\boldsymbol \Sigma}$ of the predictors set to the empirical variance-covariance matrix of 44 of the transciption factors from the real data presented in Section \ref{sec:Appli}. In this case, we make the number of non-zero parameters in vector $\bbeta^*$ vary in $p_0 \in \{5, 10, 20\}$. We compute the one-step lasso, calibrated with either the simple cross-validation scheme or the nested one, as well as the standard lasso. Results from this additional simulation study are mostly consistent with those of the previous simulation study. They confirm that calibration with the nested cross-validation scheme generally leads to better performance of the one-step lasso, in terms of support recovery (in particular precision/specificity) and prediction error. They also highlight that for low signal strengths, the proper nested scheme does not always lead to improved prediction error, especially for large $p_0$, compared to the simple (and improper) scheme. Moreover, for low signal strengths, the standard lasso performs a bit better than the one-step lasso in terms of prediction error in this setting.

\section{Application to single-cell data}\label{sec:Appli}

\begin{figure}[t]
\centering
\includegraphics[width=\textwidth]{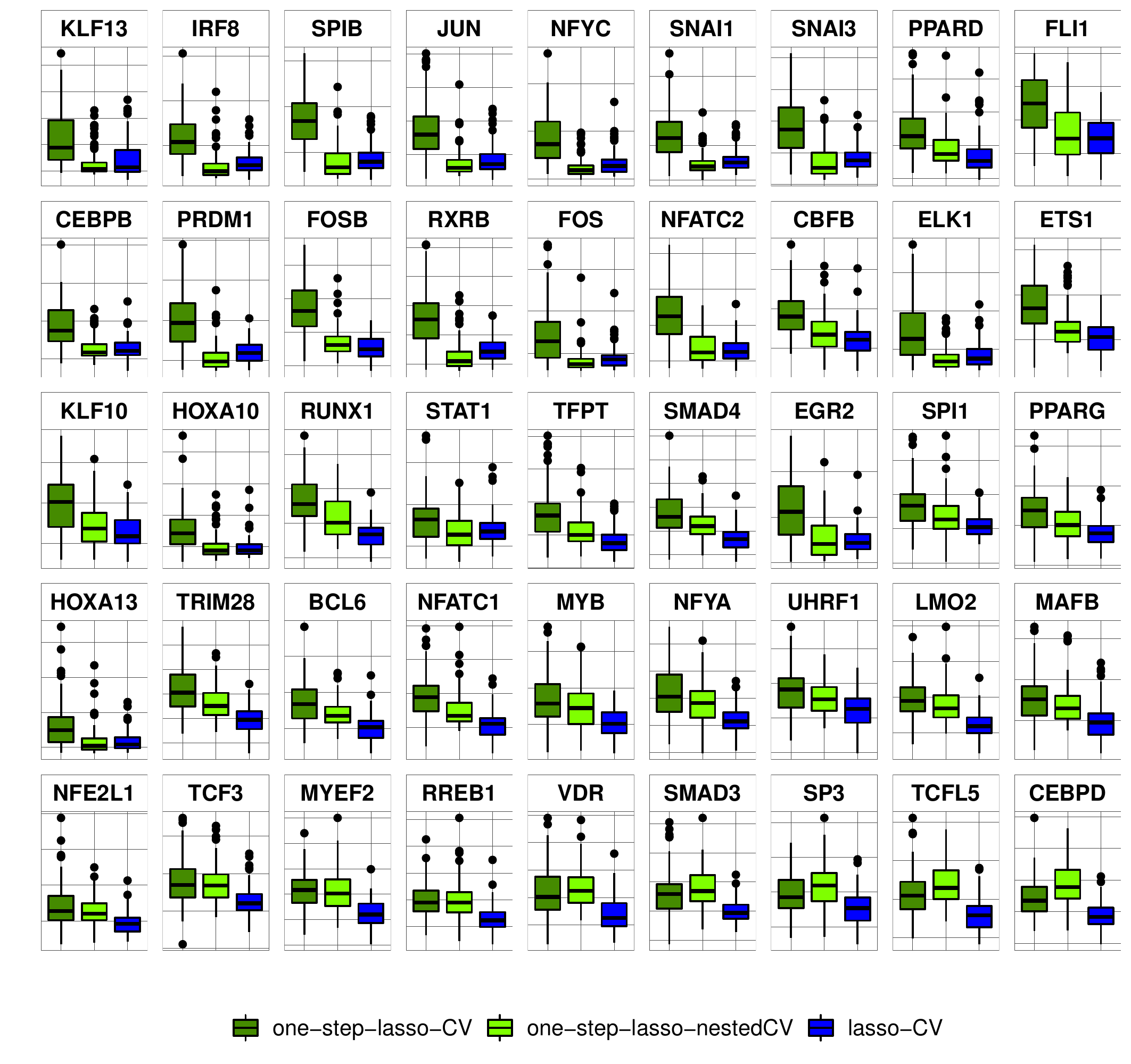}
\caption{Prediction performance of the three methods in the 45 considered linear regression models (each model corresponds to the choice of one particular transcription factor as the outcome of interest), evaluated by 5-fold cross-validation repeated 100 times. Each box-plot represents the empirical distribution of the 5-fold cross-validated prediction error over the 100 repetitions.}\label{App-PError-Zoom}
\end{figure}

In this Section, we use single-cell data to illustrate the performance of the one-step lasso calibrated by 10-fold cross-validation, using either the simple or the nested scheme. For comparison, we also implement the lasso calibrated by the simple cross-validation scheme, which is appropriate in this case. We analyse data describing mielocytic leukemia cells undergoing differentiation to macrophage, described in \cite{kouno2013temporal}; see also \cite{AutoRefLasso}. Briefly, expression levels of 45 transcription factors are measured on $n=120$ cells at 8 distinct time-points of this differentiation process. The main objective of \cite{kouno2013temporal} was to determine how associations among the 45 transcription factors vary over time. The authors focused on marginal associations and used univariate analyses, although graphical models, which describe conditional associations, might be better suited. Identification of the structure of a graphical model can be reduced to the identification and description of the neighbourhood of each covariate \citep{meinshausen2006}. For illustration, we here focus on the first time-point, and consider, successively, each of the 45 transcription factors as the outcome of interest and the other 44 factors as the predictors. This leads to the construction of 45 linear regression models, which together can be used  to construct a graphical model. Because it is out of the scope of the present work, we do not discuss, nor present, the estimated graphical models here. For each linear regression model, the prediction performance of each method (one-step lasso with the simple or nested cross-validation scheme and the standard lasso) is evaluated by 5-fold cross-validation repeated 100 times; see Figure \ref{App-PError-Zoom}. For 39 of the 45 considered models, the one-step lasso achieves a lower prediction error when calibrated using the proper nested cross-validation scheme. Moreover, even if the prediction error is generally better for the standard lasso compared to the one-step lasso (which may be explained by the low sample size in this application, and possibly low signal too), the one-step lasso outperforms the lasso for 13 models when using the proper nested cross-validation scheme for its calibration, while it never does when using the improper cross-validation scheme. Comparing the sizes of the supports for the parameter vectors produced by each method, the cumulative size of the supports for the 45 models is 464 for the lasso, 395 for the one-step lasso with improper cross-validation, and only 217 when proper cross-validation is used. Overall, these results confirm that the simple cross-validation scheme is generally not appropriate for the calibration of the one-step lasso, and that the nested cross-validation scheme usually yields sparser parameter estimates and lower prediction error.


\section{Discussion-Conclusion}\label{sec:Discussion}

In this article, we described a defect of the simple $K$-fold cross-validation scheme when applied for the calibration of the adaptive lasso, with emphasis on the ols-adaptive and ridge-adaptive lasso, as well as the one-step lasso. We further described a refined cross-validation scheme, which is appropriate for the calibration of the adaptive lasso. Although well known by specialists, we believe the defect of the simple scheme has not been stressed enough in the literature, and many statistical analysts still overlook it when calibrating the adaptive lasso. In particular, although the general principle according to which the dataset used to train the model should not contain information from hold-out sample used to evaluate prediction performance is clearly violated when using the simple scheme for the calibration of the adaptive lasso, the specific implications of applying the cross-validation after constructing the weights on the full dataset are not as straightforward to appreciate as in other situations, i.e., when applying cross-validation after screening features on the full dataset. 

We shall also stress that, although we focused on the $K$-fold cross-validation here, our description applies to other versions of cross-validation \citep{arlot2010survey}. Also, we focused on the adaptive lasso under linear regression models for simplicity, but similar defects are expected for the simple cross-validation scheme in the case of the  adaptive generalized fused lasso \citep{GenFused}, the adaptive data shared lasso \citep{Gross, AutoRefLasso, NadimSubTypes}), as well as the adaptive lasso under alternative regression models, such as generalized linear models and Cox proportional hazard models \citep{he2019improved}.

\section*{Disclaimer}
Where authors are identified as personnel of the International Agency for Research on Cancer/World Health Organization, the authors alone are responsible for the views expressed in this article and they do not necessarily represent the decisions, policy or views of the International Agency for Research on Cancer /World Health Organization.

\section*{Declarations}
\subsection*{Funding}
Not applicable
\subsection*{Conflicts of interest/Competing interests}
The authors declare no conflict of interest
\subsection*{Availability of data and material}
Not applicable
\subsection*{Code availability}
All our empirical results can be replicated using our R scripts available on \href{https://github.com/NadimBLT/AdapGlmnet-NestedCv}{https://github.com/NadimBLT/AdapGlmnet-NestedCv}. In particular the {\tt Adalasso.Nestedcv} function returns the optimal adaptive lasso estimator, with hyperparameter selected by the ``nested'' cross-validated scheme, in the case of the ols-adaptive lasso, the ridge-adaptive lasso and the one-step lasso.

\newpage
\appendix

\section{Pseudo-code of the simple $K$-fold cross-validation scheme for the calibration of the (adaptive) lasso}\label{Appendix:CV}


The pseudo-code detailed in Algorithm \ref{alg:CV} below describes the function sCVLasso: when applied to the sample $D=\{y_i,\bx_i\}_{i=1}^n$, with weights $\bw$ and number of folds $K$, it produces the adaptive Lasso estimator $\hat{\bbeta}_{\rm ada}^{\rm CV}(\bw)=\hat{\bbeta}_{\rm ada}(\lambda^{\rm CV},\bw)$, with hyperparameter $\lambda^{\rm CV}$  set to its value minimizing a $K$-fold cross-validated prediction error. Keep in mind that this simple scheme is not appropriate if weights $\bw$ are derived from the data, as is the case for the the ols-adaptive lasso, ridge-adaptive lasso and one-step lasso, estimators.

\begin{algorithm}[H]\label{alg:CV}
\SetAlgoLined
\KwData{Sample: $D=\{y_i,\bx_i\}_{i=1}^{n}$, Weights: $\bw$, Number of folds: $K$}
\KwResult{$\hat{\bbeta}_{\rm ada}^{\rm CV}(\bw)$}
 Computation of a sequence $\Lambda:= \Lambda(D, \bw) = (\lambda_1, \ldots, \lambda_{R})$\;
 Division of $D$ into $K$ folds: $D= \cup_{k=1}^K \{D^{(k)}\}$\;
 \For{$k\in\{1, \ldots, K\}$}{
 \For{$r\in \{1, \ldots, R\}$}{
   $\hat \bbeta_{r,k}$ = Lasso$(D\setminus D^{(k)}, \bw, \lambda_r)$\;
   $E_{r,k}$ = Pred.Error$(D^{(k)}, \hat \bbeta_{r,k})$\;
  }
 }
 $r^* = \argmin_r \big\{\sum_{k=1}^K E_{r,k}\big\}$\;
 $\lambda^{\rm CV} = \lambda_{r^*}$\;
 $\hat{\bbeta}_{\rm ada}^{\rm CV}(\bw)=\hat{\bbeta}_{\rm ada}(\lambda^{\rm CV},\bw)$\; 
 \caption{The simple $K$-fold cross-validation sCVLasso$(D, \bw, K)$ for the calibration the lasso. We recall that this scheme is not valid and should therefore not be used in the case of the adaptive lasso.}
\end{algorithm}

\end{document}